\RequirePackage{lineno}
\documentclass[prd,aps,preprintnumbers,amsmath,amssymb,superscriptaddress,twocolumn,floatfix]{revtex4}
\nolinenumbers
\usepackage{alltt}
\usepackage{epsfig}
\usepackage{calc,rotating,xspace}
\usepackage{pifont}
\usepackage[mediumspace,mediumqspace,Grey,squaren]{SIunits}
\usepackage{times}
\usepackage{mathptmx}
\usepackage{fancyhdr} 
\usepackage[utf8]{inputenc} 
\usepackage[T1]{fontenc}
\begin{document}
\title{\boldmath A fast method for particle tracking and triggering using small-radius silicon detectors}
\author{Ashutosh~V.~Kotwal}
\affiliation{Department of Physics, Duke University, Durham, North Carolina 27708, USA \\ ashutosh.kotwal@duke.edu}
\begin{abstract}
We propose an algorithm, deployable 
on a highly-parallelized graph computing architecture, to perform rapid reconstruction of charged-particle trajectories 
in the high energy collisions  at the Large Hadron Collider and future colliders. 
 We use software emulation to show that the algorithm can achieve an efficiency in excess of 99.95\% for reconstruction with good accuracy. The algorithm can 
 be implemented on silicon-based integrated circuits using field-programmable gate array technology. Our approach can enable a fast trigger for massive charged particles that decay invisibly in the tracking volume, as in some new-physics scenarios related to particulate dark matter. If production of 
 dark matter or other new neutral particles is mediated by metastable charged particles and is not associated with other triggerable energy deposition in the detectors, our method would be useful for triggering on the charged mediators using the small-radius silicon detectors. 
\end{abstract}
\maketitle


The reconstruction of charged-particle trajectories is one of the important experimental tasks in collider physics. Most precision measurements and searches for new physics  require reconstruction of all charged-particle trajectories with transverse momentum above a threshold. This task is usually performed with sophisticated software  which
 requires significant computing resources. This approach has been successful but faces challenges on two related fronts. Higher data rates will require corresponding increases in resources.
 Secondly, rare signals involving charged particles may be difficult to identify rapidly if they are not accompanied by other triggerable energy deposits in the detectors. Triggering 
 capability for such events has to rely on fast ``track triggers''. We propose an algorithm to perform rapid reconstruction and momentum estimation of charged-particle trajectories using the silicon detectors at small radius in 
 collider experiments. The speed is achieved by embedding the algorithm in a  highly-parallelized graph computing architecture, build using commercial field-programmable gate arrays (FPGAs).  

 Any search for new physics which involves rapid identification of high-momentum charged particles would
 benefit from our proposed method. While many final states also contain signatures detectable in the calorimeter and muon detection subsystems by existing methods, 
 a class of models predict collision events containing only ``disappearing tracks'', i.e. charged particles that decay upstream of these detector subsystems and leave
 no trace in them to enable event identification. In particular, we are interested in short-lived charged particles that travel $\cal O$(30~cm) before
 decaying invisibly. Such particles leave a trace only in the small-radius tracking detectors and disappear before even traversing 
 the large-radius tracking detectors, which extend to $\cal O$(1~m) radius. 
 In the latter case our proposed methodology would be most beneficial. 

An example is provided by models of particulate dark matter (DM) whose interaction is mediated by metastable charged particles. 
 The principles of relativistic quantum field theory and group theory, embodied in the standard model (SM) of particle physics~\cite{glashow,salam,weinberg},   have been astonishingly successful 
 in describing all known fundamental particles and their interactions including the spectacular discovery of the Higgs boson~\cite{anderson,englertBrout,peterHiggs,ghk} at the 
 Large Hadron Collider (LHC)~\cite{atlasHiggs,cmsHiggs}. 
  In parallel, the discovery of dark matter via its gravitational interactions on the galactic and cosmological distance scales~\cite{DMreview1,DMreview2,DMreview3} 
 has revolutionized our understanding of large-scale structure formation
 since the Big Bang. All cosmological data are consistent with DM comprising about 84\% of the matter in the universe~\cite{pdg}. 
 Dwarf galaxies, comprising mostly of DM, have recently been discovered~\cite{dwarfGalaxies}. 
 Since DM cannot be accounted for in the SM, it is plausible that DM
  comprises of one or more new species of particles~\cite{bertone}. Production of such particles at the LHC has been discussed in many theories of 
 particulate DM~\cite{DMcolliders}. Elastic scattering of DM particles off atomic nuclei in cryogenic materials, 
 and astrophysical detection of visible particles produced in the mutual annihilation 
 of DM particles in outer space, are being pursued by dedicated underground, ground-based and satellite-based experiments~\cite{pdg}. 

 We consider, as an example, the production of the metastable charged mediators which subsequently decay to DM particles. The known properties of the weak and electromagnetic interactions make it very 
 plausible that, if the DM particle is the lightest particle in a symmetry group multiplet, other members of the multiplet carry electric charge~\cite{neutralinos,charginos}. Therefore the LHC may 
 serve as a DM factory through copious production of the heavier, charged partners of the DM particles. 
 It is expected that the parent and daughter particles are 
 almost degenerate in mass, therefore the decay produces negligible associated energy. 
 Should the parent particles be produced in pairs, as required by known conservation laws, and they decay to DM within the volume of the experimental apparatus, the identification of such 
collision events in a time interval commensurate with the collision rate is not possible with current technology. The challenge is the identification of metastable charged mediators within the short time interval of $\cal{O}$(4$\mu$s) which is necessitated by the high collision rate, in the presence of thousands of other charged particle trajectories produced simultaneously. 

The trajectories of charged particles produced at the LHC are intercepted by highly granular silicon sensors, which at central pseudorapidities are  
 placed in concentric cylinders surrounding the colliding beams~\cite{atlasDetector,cmsDetector}. Starting in the mid-2020's, the upgraded LHC running at high luminosity (HL-LHC) will produce up to  200 proton-proton $(pp)$ collisions every 25 ns~\cite{pileup}, each collision producing about seven charged particles per unit of pseudorapidity~\cite{multiplicity}. 
 The position of each charged particle is recorded by about 10 silicon sensors along its trajectory~\cite{atlasHiggs,cmsHiggs}, resulting in a point cloud of $\approx$110,000 points recorded every 25 ns in a tracking detector covering eight units of pseudorapidity. 

Reconstruction refers to the task of partitioning the point cloud into sets of points, one point on each sensor layer, such that the elements of each set are the points created by a single charged 
 particle. Ideally, the entire point cloud is partitioned into the $\approx$11,000 sets associated with the respective particle trajectories. Current reconstruction algorithms require $\approx 4$ seconds on a single CPU core to reconstruct an event with 20 concurrent $pp$ collisions~\cite{hs06}. With an
 improved HL-LHC detector and upgraded software, the CPU time can be reduced by about a factor of two on the same CPU core, for 
 an event with 200 concurrent $pp$ collisions~\cite{hs06}. 
 A significantly faster technique,  implemented using associative memories to match the recorded patterns with stored 
 patterns, is under construction for ATLAS~\cite{FTK,FTK1}. Due to its inherent large parallelism, 
  this custom electronic device has a much shorter latency of 50-100~$\mu$s,  
 which will be reduced further to 24~$\mu$s~\cite{cerri,l1trackATLAS}. 
 However, it requires the simulation and storage of billions of patterns for matching, and is incapable of running on the $\cal{O}$(4$\mu$s) time scale.   A similar proposal~\cite{l1trackCMSam,l1trackCMSam1} has also been made for CMS. 
 In the ATLAS proposal~\cite{cerri,l1trackATLAS}, the search would be restricted to regions of interest already defined in the vicinity of other triggerable objects in the event. In another set of CMS proposals~\cite{l1trackCMS,l1trackCMS1,l1trackCMS2,l1trackCMS3,l1trackCMS4,l1trackCMS5}, a special silicon detector geometry is exploited by arranging sensor layers as closely-spaced pairs to obtain track stubs. These stubs are used as local tangents in the 
 track-finding. These stub-based approaches use the large-radius ($25 < r_{\rm sensors} < 110$~cm) silicon detector and cannot be used to trigger on 
 disappearing tracks. Techniques based on deep learning~\cite{heptrkX,heptrkX1} are applications of supervised machine learning and require training 
 data. 
 
We propose an algorithm wherein the logic is executed on a massively-parallel graph computer~\cite{graphComputing}. Since this method requires no  training on  preclassified data, it represents a type of unsupervised learning that partitions
 the input points into clusters. Compared to certain other methods of unsupervised learning, our method is not statistical and does not require large datasets to learn from, does not involve diagonalization of large
 matrices (unlike principal component analysis), and does not require initialization values (unlike k-means clustering). Furthermore, our method only uses nearest neighbor information,
 whereas k-means clustering and the expectation-maximization algorithm require a global analysis of all data points. Thus, these unsupervised learning methods cannot be processed in a parallelized, distributed computing architecture
 as our method can. We demonstrate a novel algorithm that can identify charged particle
 trajectories in a large point cloud, and how such an algorithm lends itself to implementation on a silicon integrated circuit using FPGA technology. We verify the accuracy of the algorithm and the scalability of the FPGA implementation for
 the task of triggering on metastable charged mediators. 

 Charged particles traversing an axial magnetic field in a cylindrical detector execute a helical trajectory, parameterized by functions $\phi (r)$ and $z(r)$,  where $\phi$ denotes the azimuthal   angle around the cylinder axis, and $z$ and $r$ denote the longitudinal and radial coordinates respectively. The particle's momentum perpendicular to the beam ($z$) direction,   $p_T \propto B R$ where $B$ denotes the  strength of the magnetic field and $R$ is the helix radius. Defining the curvature $c = (2R)^{-1}$ and $\phi_0$ denoting the particle's azimuthal angle at emission, we have  
\begin{eqnarray} \sin (\phi - \phi_0) = cr  \label{azimuthalTrajectory} \end{eqnarray}
If each point is recorded by a two-dimensional pixel detector, the measurement is denoted by $h$ with attributes of $\phi$ and $z$ respectively. The $z$ coordinate is given by 
\begin{eqnarray} z - z_0 = \frac{\lambda}{c} (\phi - \phi_0)  = \frac{\lambda}{c} \sin^{-1} (cr)  \label{zTrajectory} \end{eqnarray}
where the constants $\lambda$ and $z_0$ specify the cotangent of the polar angle and the $z$-position of the particle at emission, respectively. 
 It is convenient to normalize $z$ by the longitudinal length of the sensor layers in order to convert it to a dimensionless angle commensurate with the azimuthal angle $\phi$. Upon reconstruction, 
 the points associated with each particle's trajectory can be used to calculate $\lambda$, $z_0$, $\phi_0$ and $c$ (equivalently, $R$), and the latter yields a measurement of the particle's $p_T$ 
 using the knowledge of $B$. 

\section{Graph computing algorithm}
The point cloud may be represented by a two-dimensional matrix of points $h_{i,l}$ where $l$ denotes the sensor layer and $i$ denotes the point's ordinal number in that layer. For simplicity we assume 
 that all charged particles create energy deposits in all sensor layers. In future work, we will investigate relaxing this assumption so that multiple charged particles may be associated with the same 
 point in a given layer (i.e. particle trajectories may intersect at any sensor layer). Each point $h_{i,l}$ is associated with a $\phi$ measurement coordinate (in the case of one-dimensional 
 sensors) or $(\phi,z)$ measurement coordinates (in the case of two-dimensional sensors). These coordinates represent the silicon 
 pixel clusters to be provided {\em a priori} by the detector readout, similar to the scheme proposed for CMS~\cite{l1trackCMS}. 
 A cluster is defined as a group of neighbouring pixels having energy deposits above some threshold.

One may convert the $h_{i,l}$ matrix into a graph by associating links $w_{ij,l}$ between each point $h_{i,l}$ and all possible points $h_{j,l+1}$ in the adjacent outer layer of sensors. 
According to the procedure of Ref.~\cite{graphComputing} for computing derivatives on a graph, 
 the values of the link weights are set to be inversely proportional to the radial distance between consecutive layers. 
The combinatorial problem of reconstruction is equivalent to pruning the $w_{ij,l}$ matrix (i.e. eliminating all spurious links) until the surviving links partition the point cloud into sets of linked
 points and each set is associated with a physical particle trajectory. 

Viewing each point as a node in the graph, we postulate that a massively-parallel graph computer can be constructed using modern FPGA technology. Each node has a small, 
dedicated processor that stores a slice of the matrix $h_{i,l}$, and is optimized for addition, multiplication and sorting operations. 

The solution to the combinatorial problem is obtained by considering the graph operator $\Box_{ijk,l}$ at each node $(i,l)$, which is a function of first and second  
 derivatives computed using the triplet of points $h_{i,l}$, $h_{j,l+1}$ and $h_{k,l-1}$.  From Eqn.~\ref{azimuthalTrajectory} we find 
\begin{eqnarray}
\phi' & \equiv & \frac{d \phi}{dr}  =  c \sec (\phi - \phi_0) \nonumber \\
\phi'' & \equiv & \frac{d^2 \phi}{dr^2}  =  \tan (\phi - \phi_0) (\phi')^2  
\label{azimuthalLaplacian}
\end{eqnarray}
Particles of interest have high $p_T$ ($c \to 0$), implying $\phi' \to c$ and $\phi'' \to rc^3$. Therefore, $[\phi'' - r (\phi')^3] \to 0$. Similarly, 
 Eqn.~\ref{zTrajectory} yields 
\begin{eqnarray}
z' & \equiv & \frac{d z}{dr}  =  \lambda [1-(cr)^2]^{-\frac{1}{2}} \nonumber \\
z'' & \equiv & \frac{d^2 z}{dr^2}  = rc^2\lambda [1-(cr)^2]^{-\frac{3}{2}}
\label{zLaplacian}
\end{eqnarray}
Therefore $z' \to \lambda$ and $z'' \to rc^2\lambda$ for high-$p_T$ particles, and $[z'' - r (\phi')^2 z'] \to 0$. 

The graph operator $\Box_{ijk,l}$ can be computed at each node $(i,l)$ (starting at $l = 2$ because the first layer has no preceding layer to 
 compute derivatives with respect to) for all combinations of links to the previous and the next layer,
\begin{eqnarray}
\Box_{ijk,l} & = & \phi''_{ijk,l} - r_l (\bar{\phi'}_{ijk,l})^3 + z''_{ijk,l} - r_l (\bar{\phi'}_{ijk,l})^2 \bar{z'}_{ijk,l}  
\label{boxCalculation}
\end{eqnarray}
   where the first derivatives are computed as the link-weighted differences of $\phi$ or $z$ values at the two nodes connected by the link, 
\begin{alignat}{4}
& \phi'_{ij,l} && =  w_{ij,l} (\phi_{j,l+1} - \phi_{i,l}) ~ && ; ~ \phi'_{ki,l} & & =  w_{ki,l-1} (\phi_{i,l} - \phi_{k,l-1}) \nonumber \\
& z'_{ij,l} && =  w_{ij,l} (z_{j,l+1} - z_{i,l}) ~ && ; ~ z'_{ki,l} & & =  w_{ki,l-1} (z_{i,l} - z_{k,l-1}) \nonumber \\
& \bar{\phi'}_{ijk,l} && =  (\phi'_{ij,l} + \phi'_{ki,l})/2 ~ && ; ~ \bar{z'}_{ijk,l} & & = (z'_{ij,l} + z'_{ki,l})/2 
\label{derivativesCalculation}                                                                      
\end{alignat}
and the second derivatives are computed as the respective differences of 
 first derivatives at the middle (shared) node, 
\begin{eqnarray}                                                                                                                    
                                                 w_{ijk,l} & = &  2 / (w_{ij,l}^{-1} + w_{ki,l-1}^{-1}) \\
\phi''_{ijk,l}  =  w_{ijk,l} ( \phi'_{ij,l} - \phi'_{ki,l} ) & ; & z''_{ijk,l} = w_{ijk,l} ( z'_{ij,l} - z'_{ki,l} ) \nonumber 
\end{eqnarray}
 We find the criterion for valid trajectories to be $\Box_{ijk,l} \to 0$ for both one- and two-dimensional sensors, at each point of the graph. 
 In case the measurement resolutions in the two directions are unequal, Eqn.~\ref{boxCalculation} can be optimized by de-weighting the terms 
 with the worse resolution. 
 
 Eqn.~\ref{boxCalculation} encodes the list of 
 all combinatorial trajectories through node $(i,l)$. Valid local trajectories will produce small values of  $\Box_{ijk,l}$ while invalid trajectories will produce 
 large values of $\Box_{ijk,l}$. Each node is equipped with a sorting unit that ranks the values of $\Box_{ijk,l}$ over the triplets $(i,j,k)$ in ascending order. Using the sorted list of triplets, we produce 
 a sorted list of links such that the rank of a link is given by the order in which the link appears for the first time in the ordered list of triplets. The higher the rank of a link, the 
 larger the values of $\Box_{ijk,l}$ it contributes in combination with any other link.  
 Invalid links are iteratively removed by pruning the link (either $ij$ or $ik$) with the highest rank (i.e. the worst link).  
 The iterations at each node $(i,l)$ are terminated when the node is left with one link each to the next and previous layer.  

 For $N$ charged particles, the computational time cost of this algorithm involves  $\cal O$$(N)$ 
 iterations, each iteration removing the worst link at each node. Finding the worst link requires sorting $\cal O$$(N^2)$ terms. The sorting is performed once before the iterations, 
  as each iteration simply drops the last link in the list. In future work, the iterations will be optimized to prune the worst percentile in each iteration, reducing the number of iterations 
 to $\cal O$$(\log N)$.    
 Note that initially there were $\cal O$$(N^2M)$ links to be pruned, where $M$ represents the number of sensor layers. The hardware cost is the number $NM$ of graph 
 nodes, each equipped with the calculator-sorter unit. 

 Specialized algorithms and FPGA implementations for sorting large lists and finding the minimum or maximum  in a list exist~\cite{findMin1,findMin2,findMin3,findMin4}.
 Since our algorithm can proceed from course-grained to fine-grained sorting to find the best link combinations, 
 we can process $16^k$ numbers in $k$ sequential steps using a 16-input sorter.  It has been shown~\cite{Sklyarov} that a 16-input sorter for 32-bit integers 
 can be implemented on a Xilinx Spartan 6 LX45 FPGA using 5\% of its logic resources. In comparison
 to this FPGA's 43,000 logic cells, a more modern Virtex-7 2000T FPGA has 4.4 million logic cells, a factor of 100 increase in available hardware resources.
  Thus, modern FPGAs can accommodate $\cal O$(2k) 
 sorter units, which we show below to be adequate for our implementation.   
 The exponential growth of data processing worldwide has created a growing field of R\&D into FPGA-based co-processors and accelerators for data sorting and ranking, to augment traditional
 CPU-based search algorithms. Our methodology
 is well-situated to take advantage of these technical developments in FPGA sorter architectures. 

 Additional hardware will be required for reading out and
 pre-processing the raw detector hits into clusters and routing this information to the track-finding circuits. We expect these tasks to be similar to the
 readout and routing requirements of the other proposals for ATLAS and CMS track 
 triggering~\cite{FTK,FTK1,l1trackATLAS,cerri,l1trackCMS,l1trackCMSam,l1trackCMSam1,l1trackCMS1,l1trackCMS2,l1trackCMS3,l1trackCMS4,l1trackCMS5,l1track},
 and synergistic solutions for these  requirements can be pursued. 
 
\section{Results}
\begin{figure*}[h]
\begin{center}
\includegraphics*[width=8.9cm]{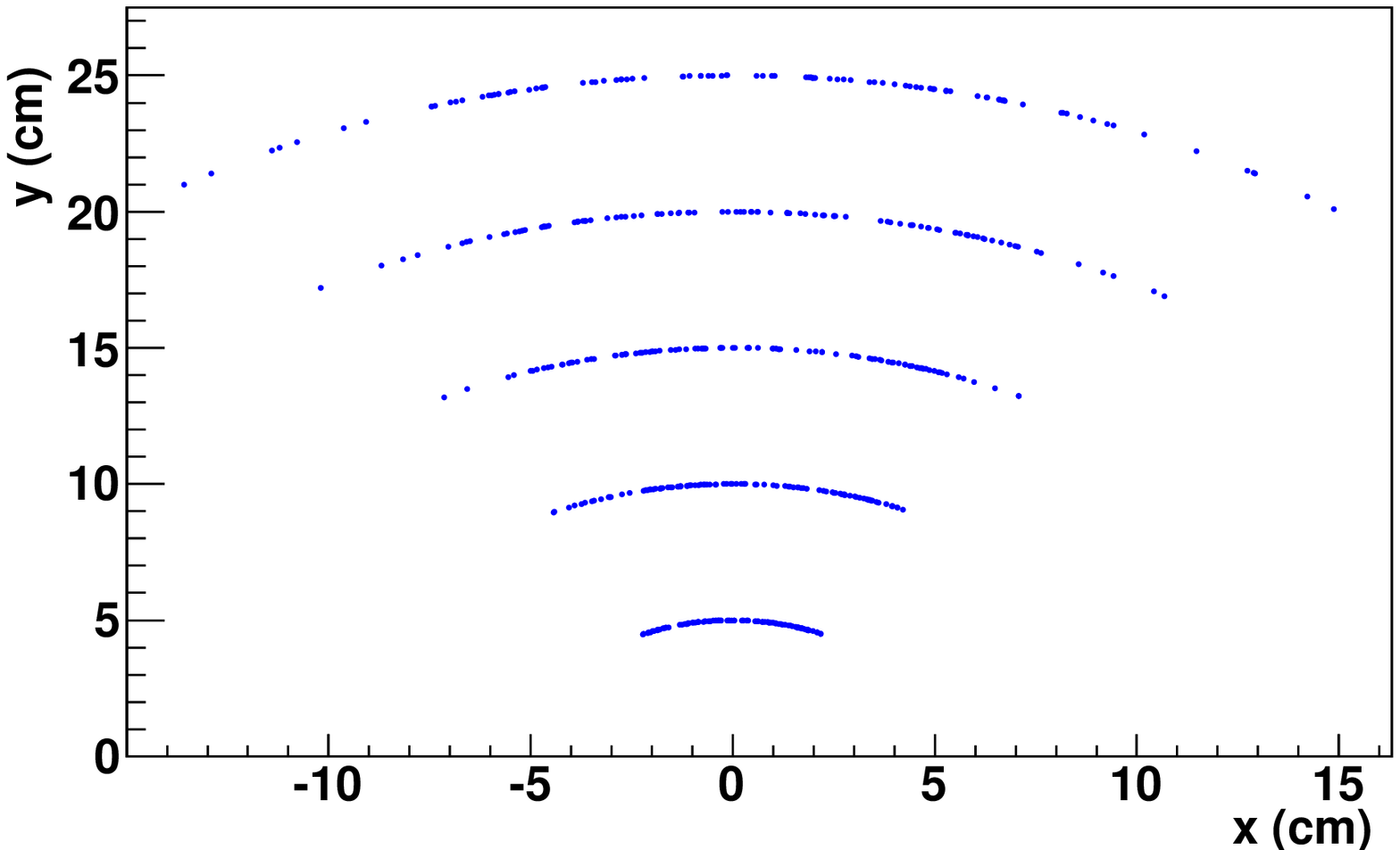}
\includegraphics*[width=8.9cm]{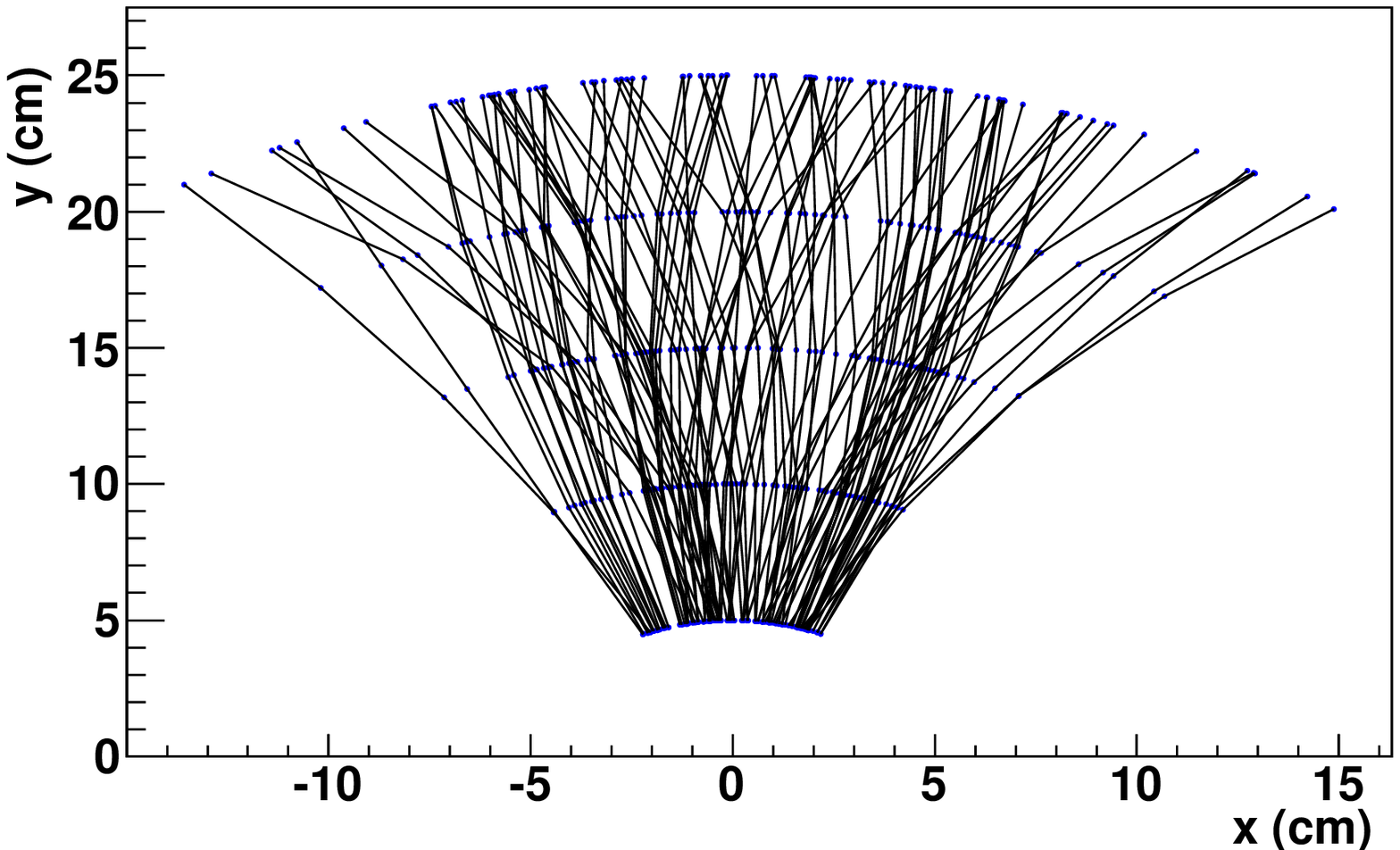}
\includegraphics*[width=8.9cm]{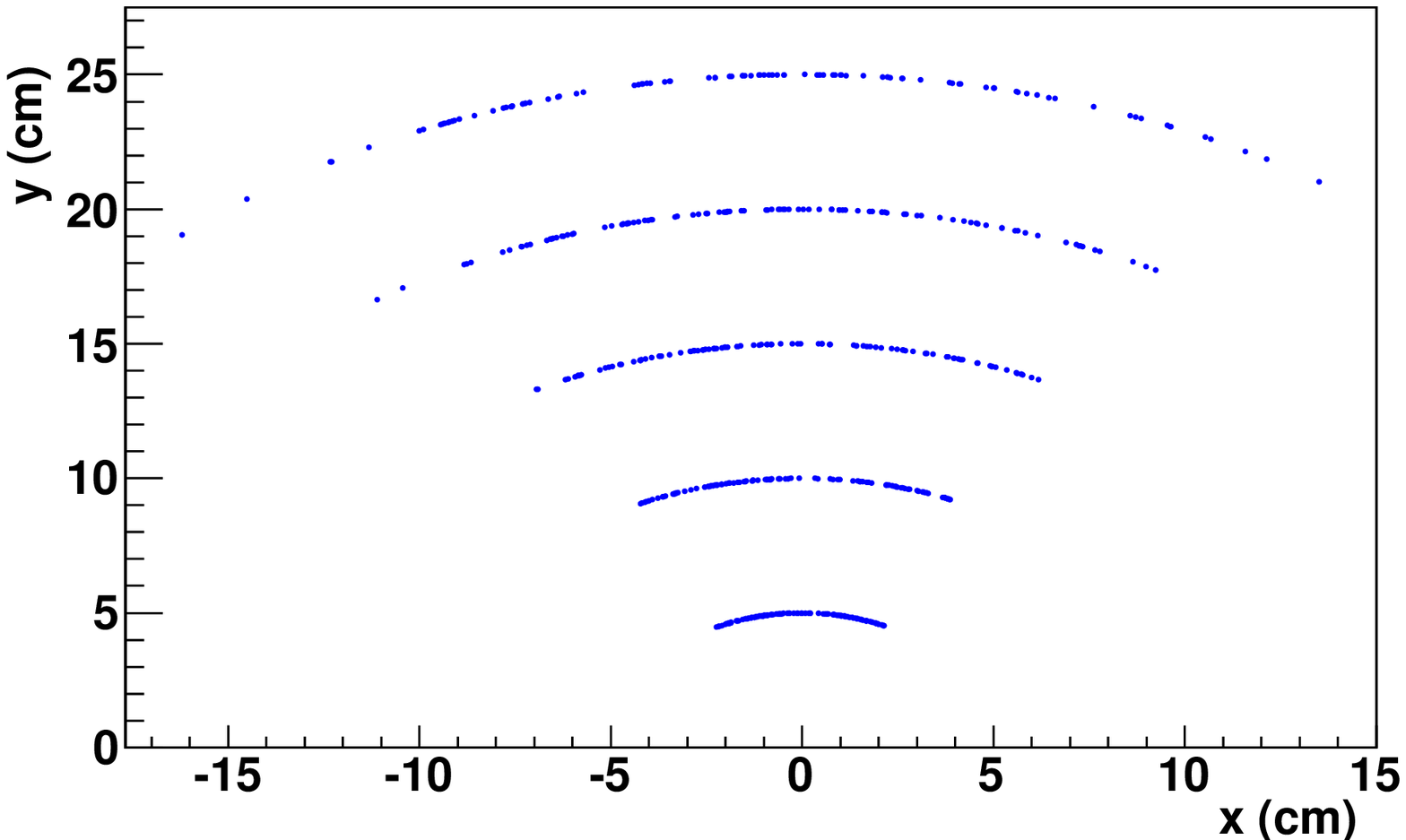}
\includegraphics*[width=8.9cm]{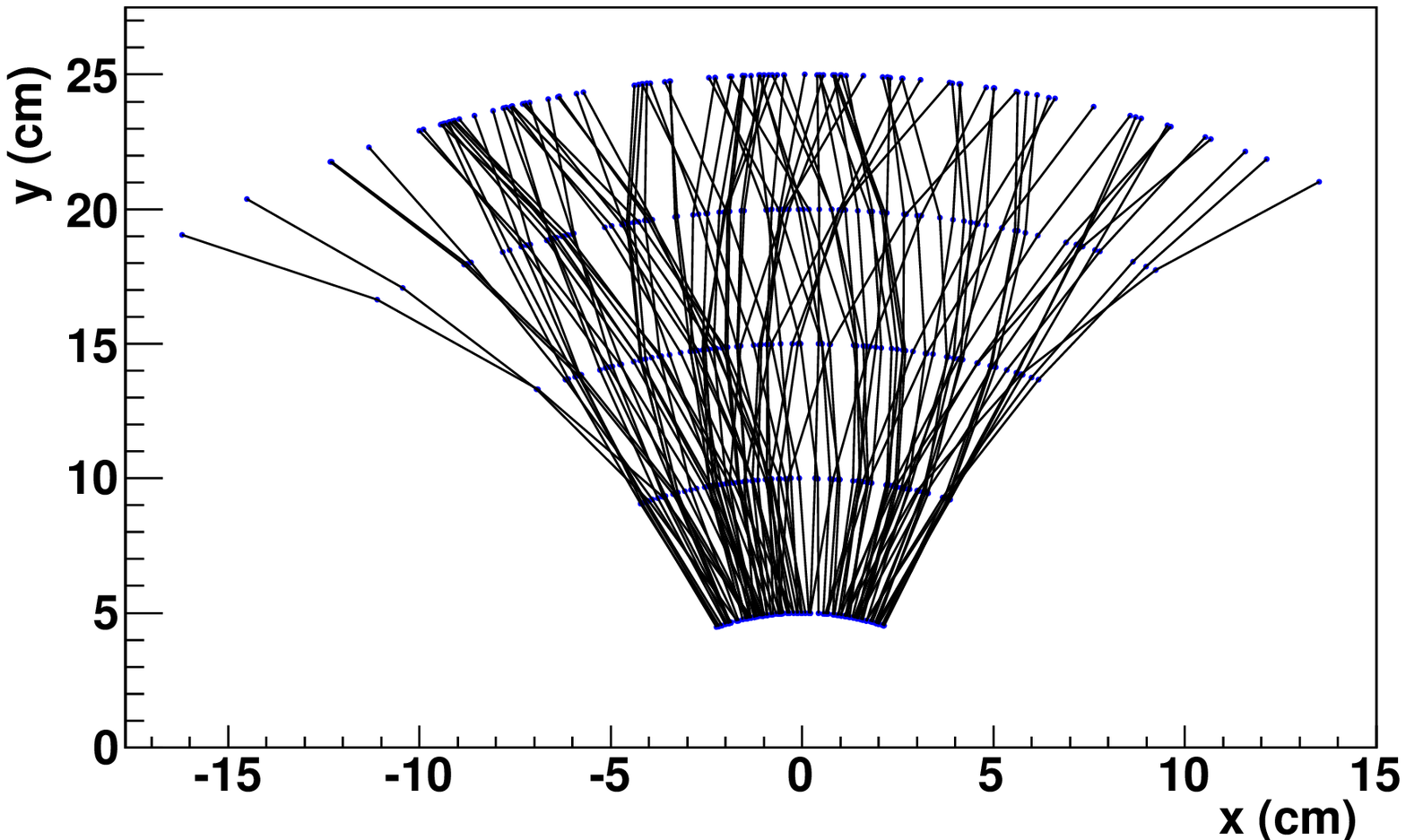}
\caption{(left) Two examples of the point cloud generated by 100 particles 
 in an azimuthal  sector of the silicon pixel detector of width one radian. The silicon sensors are 
placed in concentric circles with radial separation of 5 cm. (right) 
 The reconstruction of 100 tracks from the respective point clouds.}
\label{pointcloud}
\end{center}
\end{figure*}

 The success of this algorithm is demonstrated by the following emulation. We generate point clouds, shown in Fig.~\ref{pointcloud}, from the intersections of 100 particles traversing
 five silicon sensor layers spaced 5 cm apart in a 2 T magnetic field, over an azimuthal domain of width one radian. This detector geometry is representative of the upgraded ATLAS~\cite{atlasITK} silicon pixel barrel detector for the HL-LHC, which will be placed at the center of a cylindrical magnetic spectrometer of approximate radius 1~m. The CMS tracking volume is smaller but with a higher magnetic field of 3.8 T~\cite{cmsITK}. Since the $pp$ collision 
 region has a relatively small longitudinal length of $\approx 0.5$ cm, a projective slice of the silicon detector of  
 longitudinal width 15 cm would certainly contain a particle's complete trajectory in the $z$-direction. Therefore, a cylindrical detector of length 1.5~m (approximately the length of the HL-LHC ATLAS pixel barrel detector, covering four units of central pseudorapidity) illuminated by 5,600 particles would result in about 100
 particles contained in a wedge of azimuthal width one radian and longitudinal width of 15~cm. 
 
\begin{figure*}[h]
\begin{center}
\includegraphics*[width=8.9cm]{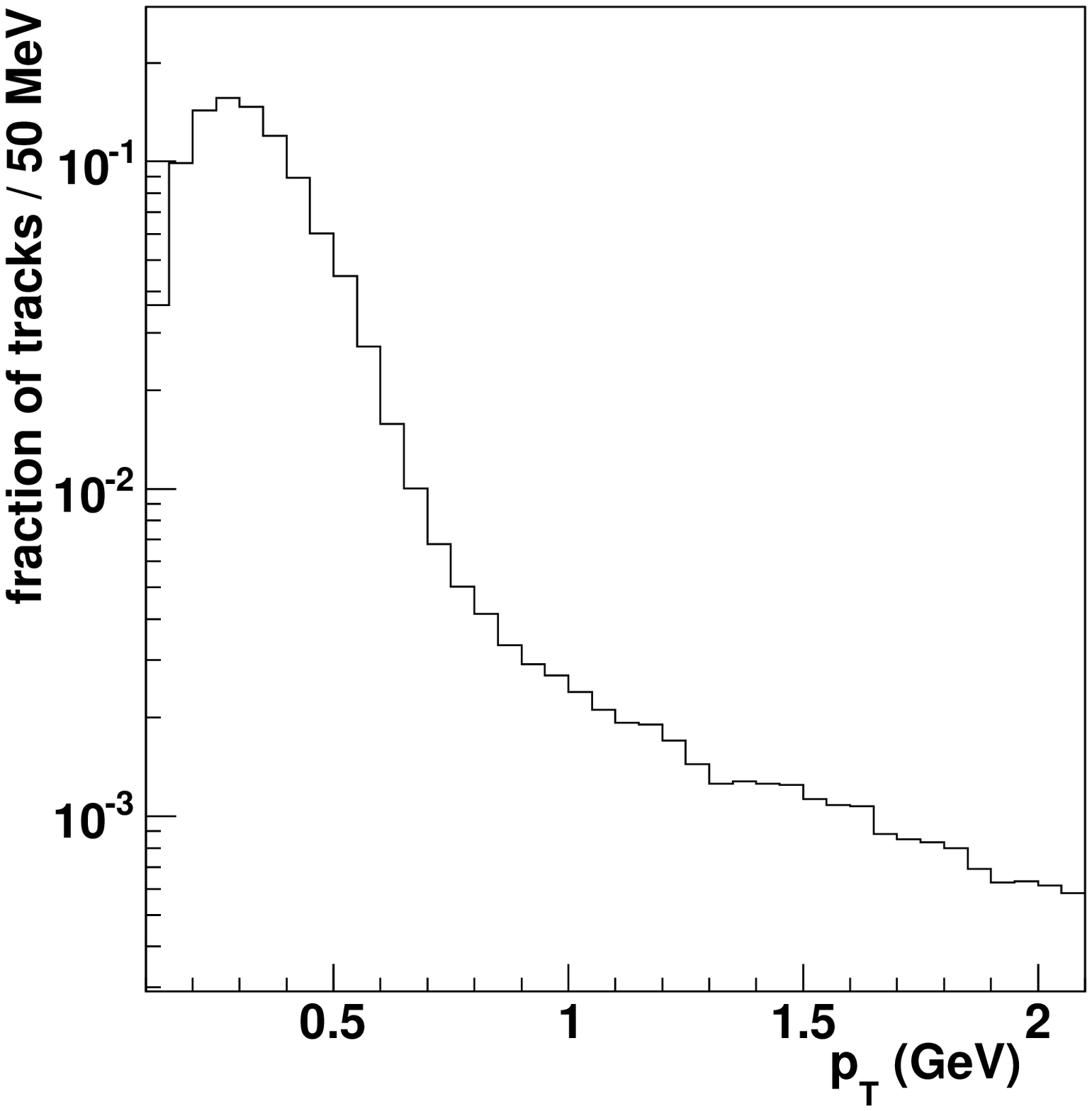}
\includegraphics*[width=8.9cm]{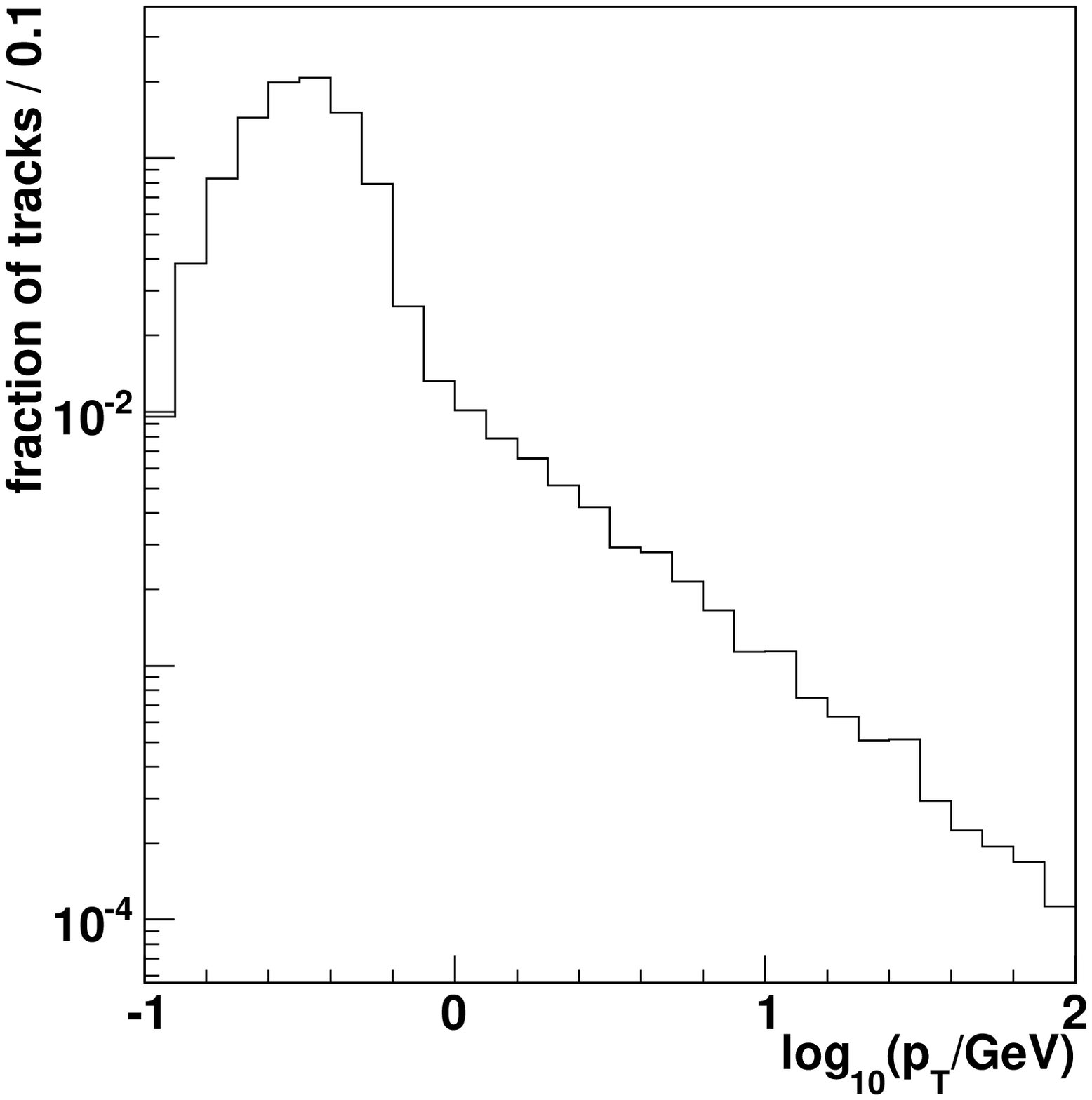}
\caption{The $p_T$ spectrum of the generated tracks used to create the point clouds of Fig.~\ref{pointcloud}.}
\label{ptDist}
\end{center}
\end{figure*}

 The transverse momentum ($p_T$) spectrum in the emulation 
 is realistically soft, modelled as an exponentially-falling distribution in $p_T^2$, $\frac{dN}{dp_T^2} \sim e^{-(p_T/\beta)^2}$. The coefficient 
 $\beta$ is chosen so that the $p_T$ spectrum peaks at $p_T = 250$~MeV~\cite{speedOfLight}, as 
 shown in Fig.~\ref{ptDist}. This model is in reasonable agreement with the measured $p_T$ spectrum~\cite{multiplicity} of soft particles 
 from minimum-bias $pp$ collisions,
 but does not provide enough statistics for studying the algorithm performance at high $p_T$. Therefore we embed a second distribution of the form 
  $\frac{dN}{dp_T} \sim p_T^{-2}$
  for $p_T > 500$~MeV, which is also soft but with a harder tail to model high-$p_T$ particles. The combined spectrum enables the study of the algorithm
 as a function of $p_T$.  

In this first study, we have not included noise hits, sensor inefficiencies and sensor resolution, in order to understand the viability
 of our approach with a perfect pixel detector. 
In subsequent studies we plan to investigate the impact of these effects on algorithm performance. 

\begin{figure}[htbp]
\begin{center}
\epsfysize = 6.07cm
\includegraphics*[width=8.92cm]{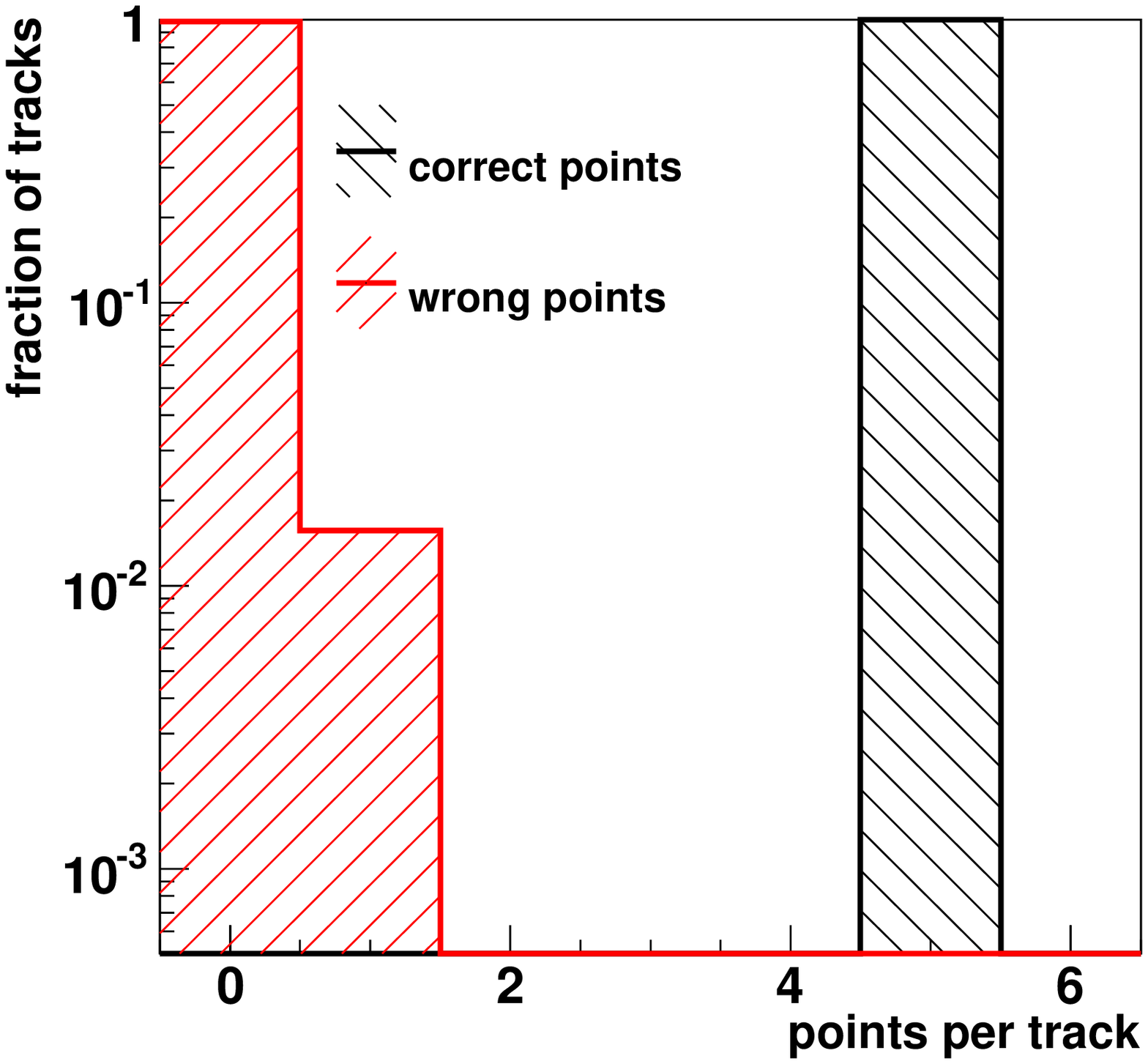}
\caption{The distributions of the number of correctly and wrongly assigned points for particle  $p_T > 500$~MeV and five sensor layers.  }
\label{efficiency1}
\end{center}
\end{figure}

\begin{figure}[htbp]
\begin{center}
\epsfysize = 6.07cm
\includegraphics*[width=8.92cm]{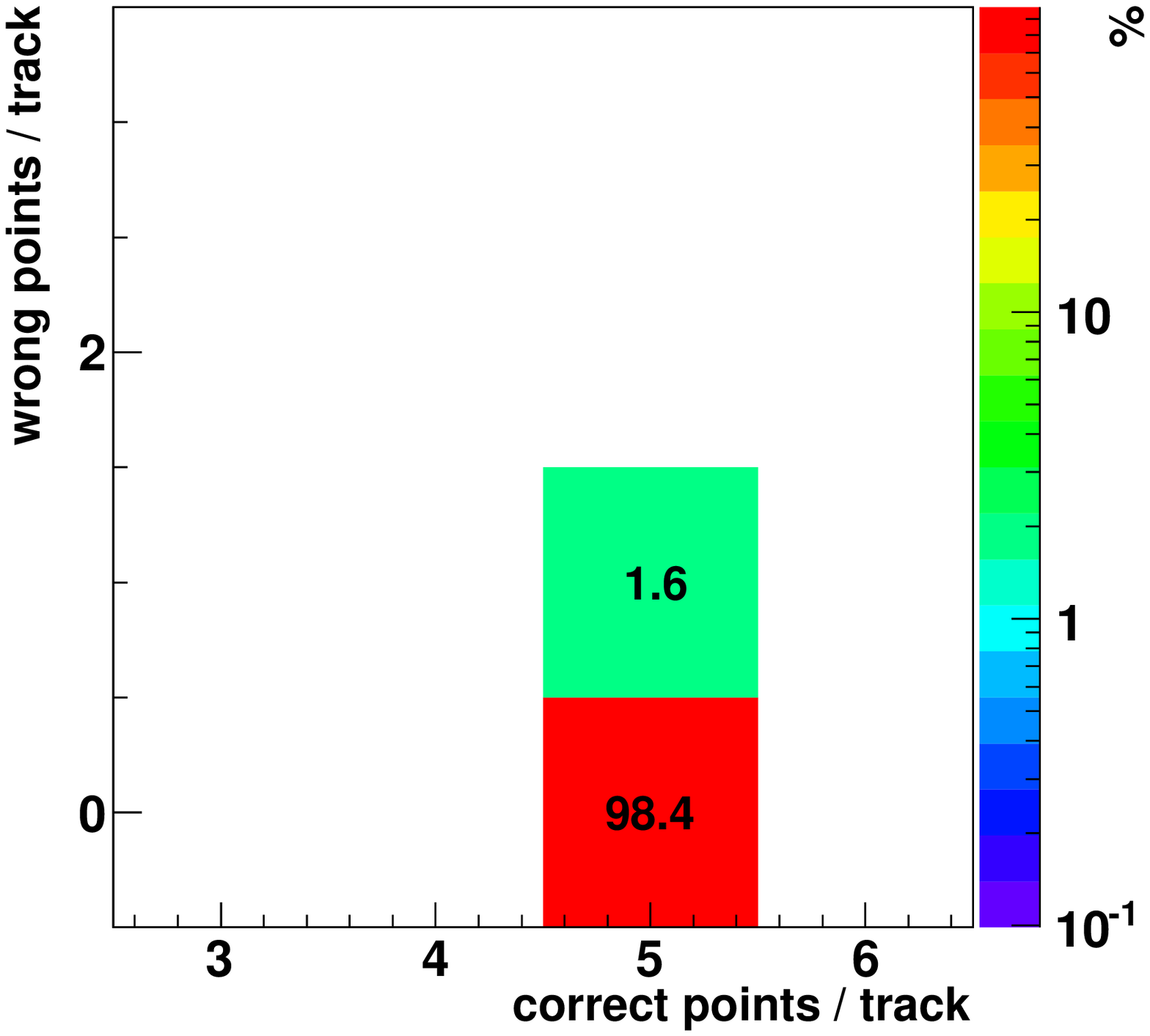}
\caption{The two-dimensional distribution of the number of correctly and wrongly assigned points, corresponding to Fig.~\ref{efficiency1}, 
 for particles with $p_T > 500$~MeV and five sensor layers.  
 The bin contents indicate the percentage of all tracks with a particular combination of correct and wrong points.  }
\label{efficiency2}
\end{center}
\end{figure}

\begin{figure}[htbp]
\begin{center}
\epsfysize = 6.07cm
\includegraphics*[width=9.3cm]{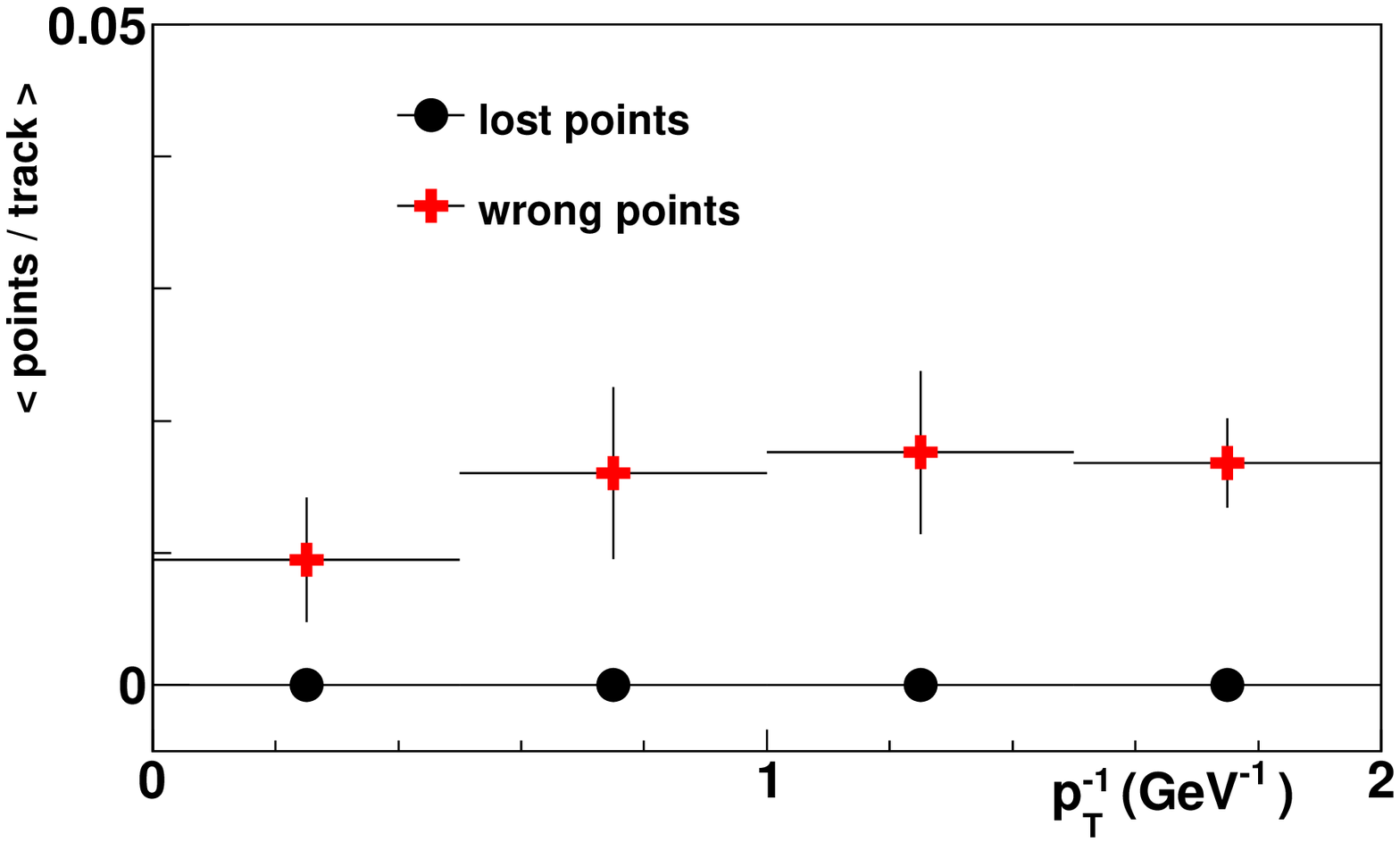}
\caption{The mean number of lost points and wrongly assigned points as a function of particle curvature (inverse transverse momentum $p_T$),
 for five sensor layers.   }
\label{effVsCurv}
\end{center}
\end{figure}

 For simplicity we show the results of the emulation using the azimuthal point coordinates only. 
 The reconstructed 
 trajectories found by the algorithm are shown in Fig.~\ref{pointcloud}. All 100 trajectories are reconstructed, and the high quality of reconstruction is demonstrated in Figs.~\ref{efficiency1}
 and~\ref{efficiency2}. Each
 point on a reconstructed trajectory is compared to its progenitor particle trajectory, and the number of correctly and wrongly assigned points per track
 are shown  in Figs.~\ref{efficiency1} and~\ref{efficiency2} respectively. 
 These rates as functions 
 of curvature are shown in Fig.~\ref{effVsCurv}. 
 These results are averaged over 160 emulated events. 

 We conclude that the track-finding efficiency exceeds 99.95\%. The probability per track to lose a correct point is $<0.1$\%
 and to have a spurious point assigned is $(1.6 \pm 0.3)$\%, with no significant momentum dependence. Thus, in the unique cases mentioned in the introduction, where new physics manifests as 
``disappearing tracks'' with no identifiable signatures at radii beyond $\approx 30$~cm, 
 our methodology can identify such tracks with high efficiency using the small-radius tracker only. The tracks will be robust, with only about one in 100 tracks having a spurious hit assigned. Since there are five sensor layers, the rate of hit loss is $<0.02$\% 
 and the rate of spurious hit assignment is $(0.32 \pm 0.06)$\%. 

\section{Discussion}

 Our algorithm can be implemented on a single FPGA integrated circuit containing an array of $100 \times 5$ arithmetic-sorter units, capable of processing
 100 tracks in a wedge with 5 points per track. The implementation is eminently possible on modern FPGAs containing
 billions of transistors. As mentioned above, a Virtex-7 2000T FPGA with its 6.8 billion transistors can accomodate $\cal O$(2k) arithmetic-sorter units, adequate for the 500 units 
 needed for this implementation. 
 Furthermore, in comparison with the 28 nm transistor feature size used in the manufacture of the Virtex-7 series, the latest Versal series of FPGAs from Xilinx are manufactured using 7 nm 
 feature size, and  contain 50 billion transistors. The steadily increasing logic resources available can only increase the processing speed and versatility of our method's implementation. 
 To process  $100^2 = 10,000$ link pairs would require running the 16-input sorters for four sequential iterations, which is acceptable from the latency
  perspective. According to Ref.~\cite{mueller}, the
 latency for a 16-input sorter is 40 ns for a 220~MHz clock frequency on a 
 Xilinx Virtex-5 FX130T FPGA, leading to an estimate of 160 ns for a progressive course- to fine-grained search for the best links. 

 Following the task of track pattern recognition, triggering schemes typically require a second step of track-fitting and momentum 
 estimation~\cite{FTK,FTK1,l1trackATLAS,cerri,l1trackCMS,l1trackCMSam,l1trackCMSam1,l1trackCMS1,l1trackCMS2,l1trackCMS3,l1trackCMS4,l1trackCMS5,l1track}.
 In our approach,
 momentum estimation is already completed since the evaluation of the derivatives in Eqn.~\ref{derivativesCalculation} 
 provides an estimate of each track's curvature. Therefore, subsequent processing steps for track-fitting and momentum estimation are not required. The 
 trigger curvature threshold could be stored in the FPGA and applied on the found tracks to produce a trigger decision directly. 

 We also note that $\cal{O}$$(120)$ such FPGA chips will be able to perform full particle reconstruction for the LHC experiments, since one FPGA is used
 to process a pseudorapidity interval of $\approx 0.4$ and 1/6 of the azimuth. Currently, traditional software codes 
 running on computer clusters are expected to require a factor of five more computing power than the budget allows~\cite{LHCcomputing}. Our approach 
 provides a promising solution to this significant problem, in addition to providing track-triggering capability. 

    The event-processing time for our FPGA circuit implementation can be reduced dramatically if the task is restricted to higher $p_T$ particles. 
 If the minimum $p_T$ of reconstructed particles was raised to 1~GeV, the width of the azimuthal wedge processed can be reduced by a factor of 10, to 
 0.1 radians,  and still contain 
 the complete trajectory of the particle. This reduces the number $N$ of tracks to be processed by a factor of 10, thereby reducing the number of combinations to be sorted by a factor of 
 100 at each processing node. The number of nodes required is also reduced by a factor of 10. Thus our algorithm and processing circuit has the flexibility to optimize speed versus $p_T$
 threshold in order to meet timing and FPGA requirements. 

 We obtain processing-time estimates using the studies in Ref.~\cite{mueller}, which were based on the Xilinx Virtex-5 FX130T FPGA. For an even-odd/bitonic merge sorting network, an 
 FPGA implementation has the number of stages $S(n) = \cal O$$(\log^2 n)$, where $n = N^2$ integers are completely sorted. In Ref.~\cite{mueller}, a latency of 100 ns was achieved for $n=64$ 32-bit integers and clock frequency 
 $f_{\rm clk} = 220$~MHz. For a triggering 
 device with a particle $p_T > 1$~GeV threshold, $N = 10, ~ n = 100$ and the latency extrapolates to 130 ns, which is significantly smaller than the 4 $\mu$s upper limit set by the LHC experiments.
 Thus, from the timing perspective, our approach is viable as a high$-p_T$ triggering device for disappearing tracks. For a pipelined implementation, the latency scales as $S(n) / f_{\rm clk}$~\cite{mueller} and can be extrapolated to other FPGA clock frequencies.  
 The speed is achieved by sorting directly in hardware and by the highly parallelized, distributive nature of this method's  computations.  

 In subsequent studies we plan to
  investigate the effects of noise hits,  sensor inefficiencies and sensor resolution. We emphasize that the 
 algorithm has no tunable or initialization parameters and requires no training, unlike other methods of supervised or unsupervised  machine learning. Our approach can be described as the partitioning of a point cloud into graphs which minimize 
 the total Dirichlet energy. This approach is viable for finding all tracks using only the small-radius silicon pixel detectors, and for fast triggering 
 on the high momentum ones, including those that decay and disappear immediately thereafter. 

\section{Acknowledgements}
We thank Raj Iyer, Henry Greenside, Alex Cloninger and Arijit Banerjee for helpful discussions. We acknowledge support from the U.S. Department of Energy, Office of High Energy Physics grant no. DE-SC0010007.


\begin{thebibliography}{99}
\bibitem{speedOfLight} We work in units where speed of light is set to unity, thereby equivalencing the units of momentum and energy.
\end{thebibliography}
\end{document}